\documentclass[a4paper,twoside,10pt]{letter}
\usepackage{graphicx,saj,multicol,subeqnarray}
\usepackage{multirow}

\def\papertype{Original Scientific Paper}
\newcommand{\D}{$^\circ$}
\newcommand{\ESO }{\mbox {ESO\,295-IG022}}
\newcommand{\ESOs }{\mbox {ESO\,295-IG022-S}}
\newcommand{\ESOn }{\mbox {ESO\,295-IG022-N}}

\newcommand{\AB }{\mbox {Abell\,S0102}}
\newcommand{\etal}{et al.}

\newcommand{\ie}{i.e.}

\def\p0{\phantom{0}}
\def\arcmin{\hbox{$^{\prime}$}}
\def\arcsec{\hbox{$^{\prime\prime}$}}

\def\udc{...}
\begin{document}
\baselineskip=3.1truemm
\columnsep=.5truecm
\newenvironment{lefteqnarray}{\arraycolsep=0pt\begin{eqnarray}}
{\end{eqnarray}\protect\aftergroup\ignorespaces}
\newenvironment{lefteqnarray*}{\arraycolsep=0pt\begin{eqnarray*}}
{\end{eqnarray*}\protect\aftergroup\ignorespaces}
\newenvironment{leftsubeqnarray}{\arraycolsep=0pt\begin{subeqnarray}}
{\end{subeqnarray}\protect\aftergroup\ignorespaces}
%


\markboth{\eightrm Radio-continuum jets around the peculiar galaxy pair
      ESO\,295-IG022}
{\eightrm M.~D.~Filipovi\'c, E.~J.~Crawford, P.~A.~Jones and G.~L.~White}

{\ }

\publ

\type

{\ }


\title{Radio-continuum jets around the peculiar galaxy pair
      ESO\,295-IG022}


\authors{M.~D.~Filipovi\'c$^1$, E.~J.~Crawford$^1$, P.~A.~Jones$^{2,3}$ and G.~L.~White$^1$}

\vskip3mm


\address{$^1$University of Western Sydney, Locked Bag 1797, Penrith South DC, NSW 1797, Australia}

\Email{m.filipovic}{uws.edu.au}
\vspace*{-1ex}
\Email{e.crawford}{uws.edu.au}

\vspace*{-1ex}

\address{$^2$School of Physics, University of New South Wales, Sydney, NSW 2052, Australia}
\vspace*{-2ex}
\address{$^3$ Departamento de Astronom\'\i a, Universidad de Chile, Casilla 36-D, Santiago, Chile}
\Email{pjones}{phys.unsw.edu.au}


\dates{September XX, 2010}{October XX, 2010}


\summary{We report new radio-continuum observations with the Australia Telescope Compact Array (ATCA) of the region surrounding the peculiar galaxy pair \ESO, at the centre of the poor cluster \AB. We observed this cluster at wavelengths of $\lambda$=20/13 and 6/3~cm with the ATCA 6~km array. With these configurations, we achieved a resolution of $\sim$2\arcsec\ at 3~cm which is sufficient to resolve the jet-like structure of $\sim$3\arcmin\ length detected at 20~cm. From our new high resolution images at 6 and 3~cm we confirm the presence of a double jet structure, most likely originating from the northern galaxy (\ESOn), bent and twisted towards the south. We found the spectral index of the jet to be very steep ($\alpha$=--1.32). No point source was detected that could be associated with the core of \ESOn. On the other hand, \ESOs\ does not show any jet structure, but does show a point radio source. This source has variable flux and spectral index, and appears to be superposed on the line-of-sight of the jets (seen at 20-cm) originating from the northern galaxy \ESOn. Finally, regions of very high and somewhat well ordered polarisation were detected at the level of 70\%.}


\clearpage
\begin{multicols}{2}
{


\section{1. INTRODUCTION}

Merging and interacting galaxies are often found in galaxy clusters. Such interactions and mergers provide a valuable insight into the evolution of both the cluster and the embedded galaxies, thus underpinning the most current theories of cluster and galaxy formation and evolution. As an example, it is now well established that radio jets interact with the intracluster medium (ICM) providing energy to the ICM.  This is shown by Rafferty \etal\ 2006) and B\^irzan \etal\ (2007) that the synchrotron ages of the radio jet lobes are generally less than the X-ray ages of the cluster cavities; suggesting that the cavities are being pumped by the energy supplied by the jet. In their study, they compared the size of the cavities (determined from X-ray data) with the size of the lobes (found on radio maps) and concluded that the lower radio frequencies maps are a better tool (than the higher frequency maps) the several hundred million year history of active clusters.

A prime example of a complex, merging and interacting system galaxies within a cluster has been discovered by Read \etal\ (2001). They reported a prominent (bipolar) radio source with jets embedded in the poor galaxy cluster \AB\ (Abell \etal\ 1989) (also known as EDCC~494; Lumsden \etal\ 1992) at a redshift of z=0.054824 (Bica \etal\ 1991). The \ESO\ galaxy pair also coincides with \AB\ and was classified as a merging galaxy system by Bica \etal\ (1991). 

Read \etal\ (2001) suggested two possible scenarios for the appearance of radio-continuum jet/s. One is a ``single galaxy'' model with two jets coming out of either \ESOs\ or \ESOn\ and with the other galaxy seen superimposed on the jet structure. For this scenario they suggest the southern galaxy (\ESOs) as the ``origin'' of the bipolar radio jets which extend south about 95\arcsec\ \mbox{($\approx100$\,kpc} at the distance of \ESO), and to the north about $\approx80$\arcsec.  The northern termination is positional coincidence with the northern galaxy \ESOn. 

The alternatively scenario suggested by Read \etal\ (2001) is that the emission emanates from the northern galaxy \ESOn\ and that there is a positional coincidence with a similar jet structure associated with \ESOs. This second, and less likely scenario, requires the northern and southern galaxies to produce their own substantive jets.  In this two jet model, Read \etal\ (2001) estimate projected lengths of up to 100\,kpc, with velocities of at least 1000\,km s$^{-1}$, and explain the bending of the jets as being due to \ESOs\ moving through the ICM at \mbox{$\sim190$\,km s$^{-1}$}. 

The ROSAT PSPC observations of Read \etal\ (2001) indicate relatively cool diffuse X-ray emission from \AB\ consistent with group or poor cluster emission. This emission, when compared with the radio jets, is suggestive of channeling effects taking place that might create the so-called cavities where jets are able to punch holes in the ICM and displace the X-ray emitting gas, as is seen in other galaxy groups/clusters such as Hydra~A (Wise \etal\ 2007), Abell\,2204 (Sanders \etal\ 2009) or Abell\,1446 (Douglass \etal\ 2008). 

Our aim was to use Australia Telescope Compact Array (ATCA) high-resolution radio-continuum observations to resolve the jet structure of this Abell cluster, and to better determine the relationship between the optical galaxies and radio-continuum/X-ray (anti)correlation. In this paper we present new ATCA radio results of this cluster area at 6 and 3~cm. In Sect.\,2, we describe our radio-continuum observations and data analysis. The results of this analysis are given in Sect.\,3. Finally, we summarise our findings in Sect.\,4.

\section{2. OBSERVATIONS AND DATA ANALYSIS}

The \AB\ region was initially observed as part of the ATCA observations of the NGC~300 area (ATCA project C828) at wavelengths of 20 and 13~cm ($\nu$=1384 and 2496~MHz) with a 6A array giving an angular resolutions of 6$\arcsec$ and 4$\arcsec$ (see Figs.~1 and 2). More information regarding these observations can be found in Payne \etal\ (2004). As the \AB\ positions are well down the primary beam pattern of the ATCA, we have applied a primary beam correction using the standard techniques in the {\sc miriad} software package (Sault \& Killeen 2010).

We again observed \AB\ with the ATCA on 9$^\mathrm{th}$ October 2000 (ATCA project C913), with an array configuration 6A, at wavelengths of 6 and 3~cm ($\nu$=4800 and 8640~MHz). These observations were undertaken in so-called full synthesis mode and total $\sim$10.5 hours of integration over a 12 hour period. Source 1934-638 was used for primary calibration and source 0048-427 was used for secondary calibration. The \textsc{miriad} (Sault \& Killeen 2010) and \textsc{karma} (Gooch~2006) software packages were used for data reduction and analysis.

The 6\,cm image (Figs.~2 and 3) has a resolution of 4\arcsec$\times$3\arcsec\ at position angle 2\D\ and the r.m.s noise is estimated to be 0.1~mJy/beam. Similarly, the 3\,cm image was constructed with a resolution of 2.2\arcsec$\times$1.6\arcsec\ and r.m.s noise of 0.1~mJy/beam (Fig.~4).

Additional archival $\lambda=6$\,cm ATCA data from 13$^\mathrm{th}$ December 1991 (ATCA project C133), centred on \ESO were found. These observations are only $\sim$20 minutes of fairly low sensitivity observations, and were only to determine \ESOs\ flux and position. Finally, we note recent (17$^\mathrm{th}$ January 2010), Burnett \etal\ (2010) ATCA CABB observations with some significant calibration problems, from which they managed to extract integrated flux densities for the main feature of \ESOs\.  We report their results in our Table~1 (also see Fig.~5c).

Two prominent radio features are seen at 13 and 6~cm -- the ``core/s'' and ``jet/s''. The radio point source (J005546-372427: at RA(J2000)=00$^{\rm h}$55$^{\rm m}$46.58$^{\rm s}$, {DEC(J2000)=- 37$^{\circ}$24\arcmin27.7\arcsec)} is coincident with the southern galaxy (\ESOs). The ``jet'' -- named J005547-372320 (at RA(J2000)=00$^{\rm h}$55$^{\rm m}$47.31$^{\rm s}$, DEC(J2000)=--37$^{\circ}$23\arcmin20.6\arcsec) appears to be associated with optical counterpart \ESOn (positions from the 6~cm observations). 

The frequency-dependent integrated flux densities ($S$) of J005547-372320 and J005546-372427 features are given in Table~1 and shown in Fig.~5. We point out that we do not detect the jets at 3~cm because of the very steep radio spectrum of the jets, however, this steep spectral index  ($\alpha=-1.32\pm0.14$; see Fig.~5a) may be the result of the missing flux of the short spacings at higher radio-continuum frequencies (6 and 3~cm).

The integrated flux density of J005547-372320 was determined as the sum of the flux density in a box around \ESOn\ and that of the two twisted jet features. The integrated flux density of the central point-like radio source coincident with \ESOs\ was determined using the two-dimensional elliptical Gaussian fitting algorithm within the \mbox{MIRIAD} software package. The spectral index distribution of this object is shown in Figs.~5b and 5c. We emphasise that the integrated flux at 20~cm of the \ESO\ source in our 28$^\mathrm{th}$ February 2000 observations agrees well with the flux density from the VLA NVSS and that reported by Read \etal\ (2001). We estimate the integrated flux density errors from all our images to be less than 10\%.

}\end{multicols}


%
%
\centerline{\includegraphics[width=\textwidth]{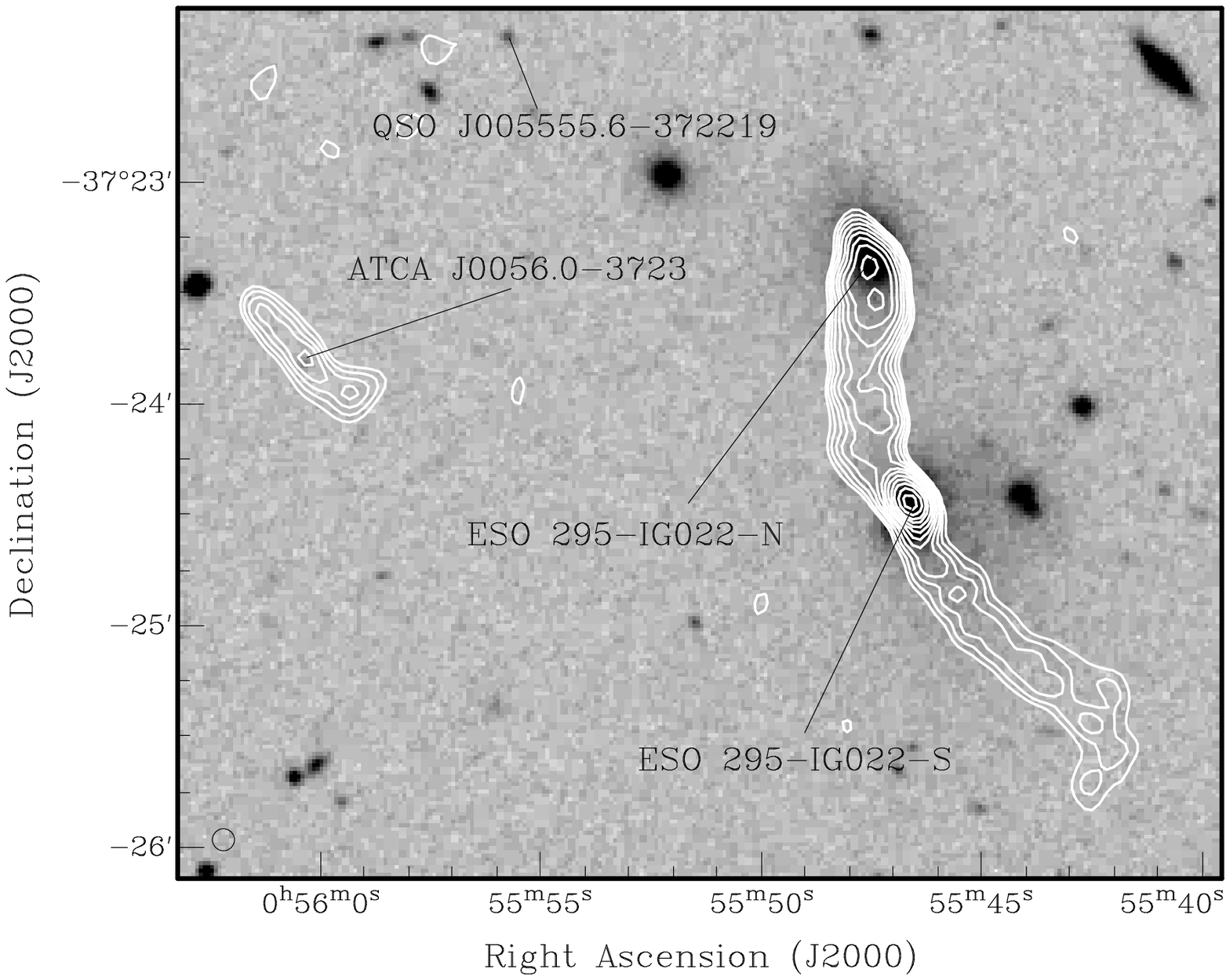}}
\vspace*{-0.5cm}
\figurecaption{1.}{ Digital Sky Survey (red) image of the Abell~S0102 cluster central region overlaid with primary beam corrected 20~cm ATCA radio-continuum contours. The synthesised beam of the ATCA observation is 6$\arcsec$ $\times$ 6$\arcsec$ (see circle, lower left corner) with r.m.s. noise (1$\sigma$) of 0.06~mJy. Contours increase by factors of $\sqrt{2}$ from 0.5~mJy/beam. This picture is adapted from Read \etal\ (2001).}


{\center\textbf{Table 1.} Radio-continuum integrated flux densities ($S$) of the two radio features within Abell\,S0102. Our new spectral index ($\alpha$) estimates are based on two observing sessions from 2000 while Burnett \etal\ (2010) is also shown in Column~8. }

\scriptsize
 \noindent \begin{tabular}{lllcccccl}
\hline
\noalign{\smallskip}
ATCA                  & Other & Obs. & $\rm S_{\rm 20\,cm}$ & $\rm S_{\rm 13\,cm}$ & $\rm S_{\rm 6\,cm}$ & $\rm S_{\rm 3\,cm}$& $\alpha \pm \Delta \alpha$ & Reference \\
Radio Name            & Name  &  Date     &  (mJy)   & (mJy)  & (mJy)  & (mJy)  & & \\
\noalign{\smallskip}
\hline
\noalign{\smallskip}
J005547-372320  & \ESOn & 28/02/2000 &    80.0  & 42.5  &       & --     & \multirow{2}{*}{$-1.32\pm0.14$}  & Read \etal\ (2001)\\
J005547-372320  & \ESOn & 09/10/2000 &          &       & 15.5  & --     & & This paper\\
\noalign{\smallskip}
J005546-372427  & \ESOs & 13/12/1991 &          &       & 32.6  &        & & Read \etal\ (2001) \\
J005546-372427  & \ESOs & 28/02/2000 &    47.0  & 63.9  &       &        & \multirow{2}{*}{$-0.11\pm0.14$}  & Read \etal\ (2001)\\
J005546-372427  & \ESOs & 09/10/2000 &          &       & 48.6  & 41.1   & & This paper\\
J005546-372427  & \ESOs & 17/01/2010 &    18.2  &       & 74.0  & 67.0   & $+0.78\pm0.31$ & Burnett \etal\ (2010)\\
\noalign{\smallskip}
\hline
  \end{tabular}
  
  \normalsize


\begin{multicols}{2}
{

\section {3. RESULTS AND DISCUSSION}

In comparing our various radio-continuum images with optical Digital Sky Survey~2 (DSS2) images (Figs.~1, 2, 3 and 4), we see some striking features in and around the central \AB\ cluster galaxy pair, \ESO. From the higher resolution image at 13 and 6~cm (Figs.~2 and 3), the southern peak appears quite compact and NOT a knot in the jet. The compact nature of this source is supported by our 2000 observations of \ESOs\ that give a flat spectrum of $\alpha=-0.11\pm0.14$ (Fig.~5b). This spectral index is, however, significantly different from $\alpha=+0.78\pm0.31$ found in 2010 by Burnett \etal\ (2010) (Fig.~5c) and we note that the flux density values for Burnett \etal\ (2010) are higher than ours by a factor of ~1.6 at the higher frequencies (6 and 3 cm) but substantively lower than that of Read \etal\ (2000) by a factor of ~0.4 at the lower (20 cm) frequency. We are, therefore, assuming that the source is both of flat spectrum and variable, and interpret the inconsistencies of flux density as the consequence of either variability (over two decades) or the loss of flux due to the missing spacings in the data set/s.

As pointed by Read \etal\ (2001), the galaxy \ESOs\ has a complex optical morphology with two nuclei $\approx6$\arcsec\ (\ie\ 6.5\,kpc) apart. This may be a merging pair or a NGC 5128 type object seen at great distance. The bright and compact northern-most radio knot at 3~cm (RA(J2000)=00$^h$55$^m$46.57$^s$, DEC(J2000)=$-$37\D24\arcmin27.5\arcsec) lies within $\sim1$\arcsec\ of the northern component of \ESOs\ (Fig.~4) and, given the positional errors of both the optical and radio data, (both $\approx$1\arcsec), it seems sensible to interpret the point radio source as coincident with \ESOs and emanating from the AGN core of the northern component of that galaxy. 

The scenario put forward by Read et al (2000), that \ESOs\ is the origin of the observed (20~cm) jets, is therefore unlikely. We do not find the jet structure asymmetric as the northern jet is much brighter than the southern. Combining this with the compact and potentially variable nature of the radio source and its identification with a component of \ESOs. We argue that the jet structure most likely originates from \ESOn\ and not from \ESOs\ as suggested by Read et al (2000). This is therefore probably a rare, but not too unlikely random line of sight coincidence.

%
%
\centerline{\includegraphics[width=9cm,angle=-90]{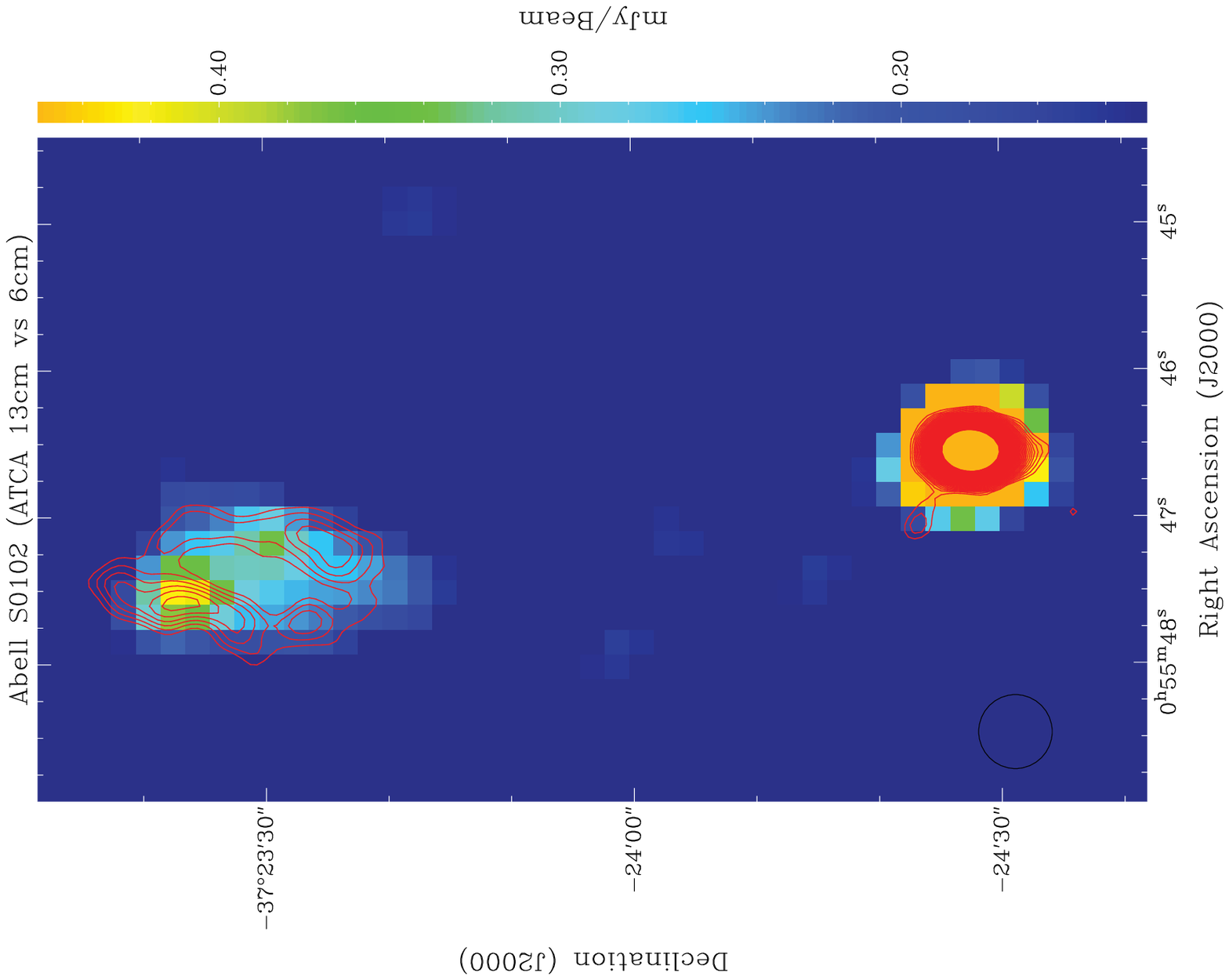}}
\vskip-3mm
\figurecaption{2.}{ATCA 13-cm image of the Abell~S0102 cluster central region overlaid with primary beam corrected 6-cm ATCA radio-continuum contours. The synthesized beam of the ATCA observation is 4$\arcsec$ $\times$ 3$\arcsec$ (see circle, lower left corner) with r.m.s. noise (1$\sigma$) of 0.1~mJy. Contours increase by factors of 2.5$\sigma$ from 0.5 mJy/beam.}

%
%
\centerline{\includegraphics[width=8.5cm]{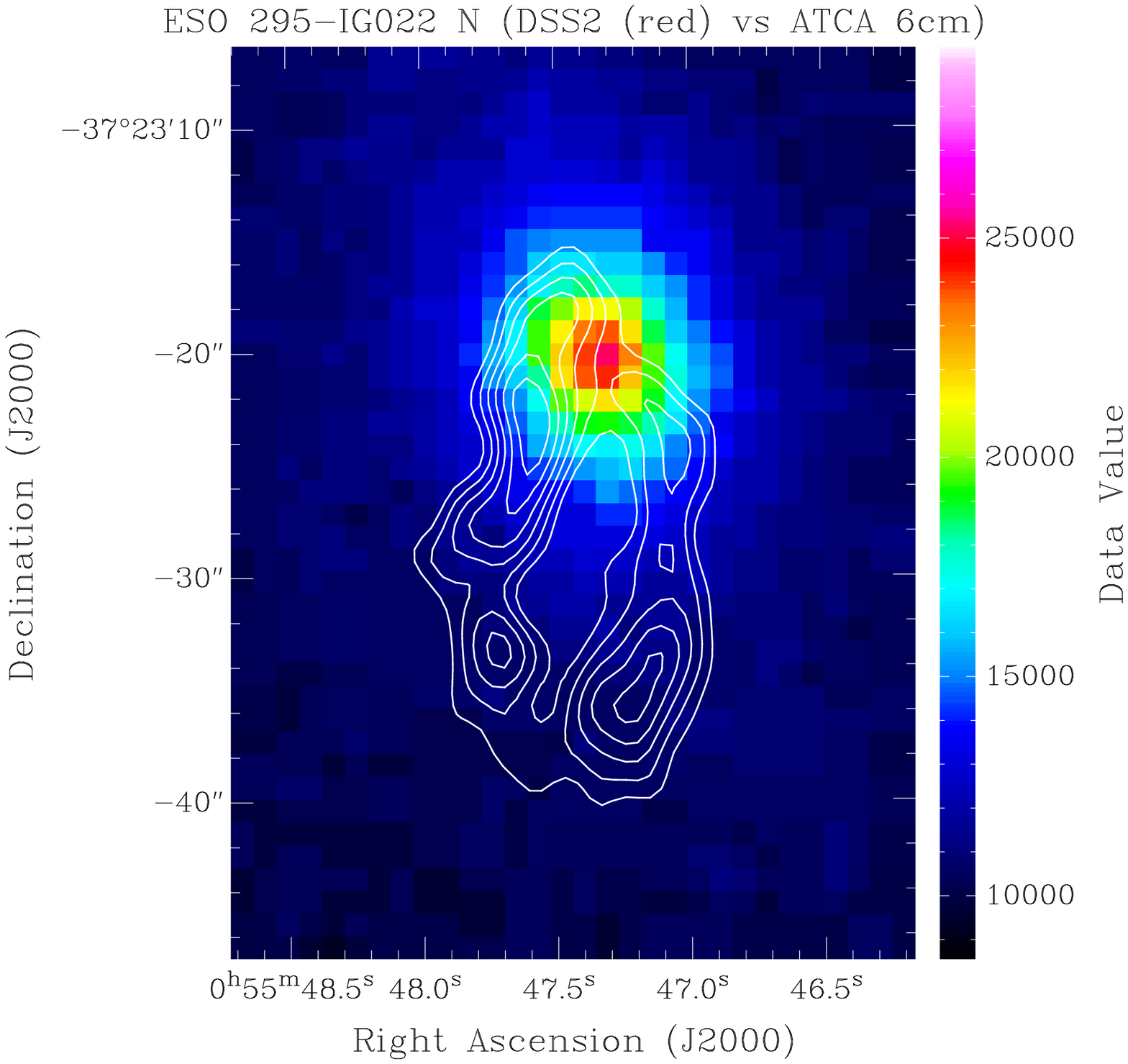}}
\vskip-4mm
\figurecaption{3.}{Digital Sky Survey (red) image of the \ESOn\ region overlaid with primary beam corrected 6-cm ATCA radio-continuum contours. Contours are the same as in Fig.~2.}

The 6~cm twin jet of the northern source (Fig.~3) is nicely coincident with a optical object which is an extended and elliptical galaxy. We point out the lack of a point radio source which could be associated directly to \ESOn\ but see the twin-jet nature of the jet near \ESOn\ as a defining feature (this is not seen in the 13~cm image with lower resolution). As radio cores are often much weaker than the jets, the non detection of a core in \ESOn\ is not too surprising (see Jones \etal\ 1994; cores typically $<$1\%\ of jets). Finally, our spectral index estimate for this jet feature is quite steep  -- $\alpha=-1.32\pm0.14$ (Fig.~5a). This steep spectral index may be an explanation for not detecting the jet features in our 3~cm image. However, the steep spectral index is, at least in part, due to missing flux at higher resolution/frequency. McAdam \etal\ (1988) reported discovery of radio source PKS\,0427-53 interacting with IC\,2082 and the cluster medium. The morphology of this radio AGN looks very similar to \ESOs. Even the size of PKS\,0427-53 jets of 160~kpc is similar to \ESOs\ jets estimate of 180~kpc.

%
%
\vskip-8mm
\centerline{\includegraphics[width=8.5cm]{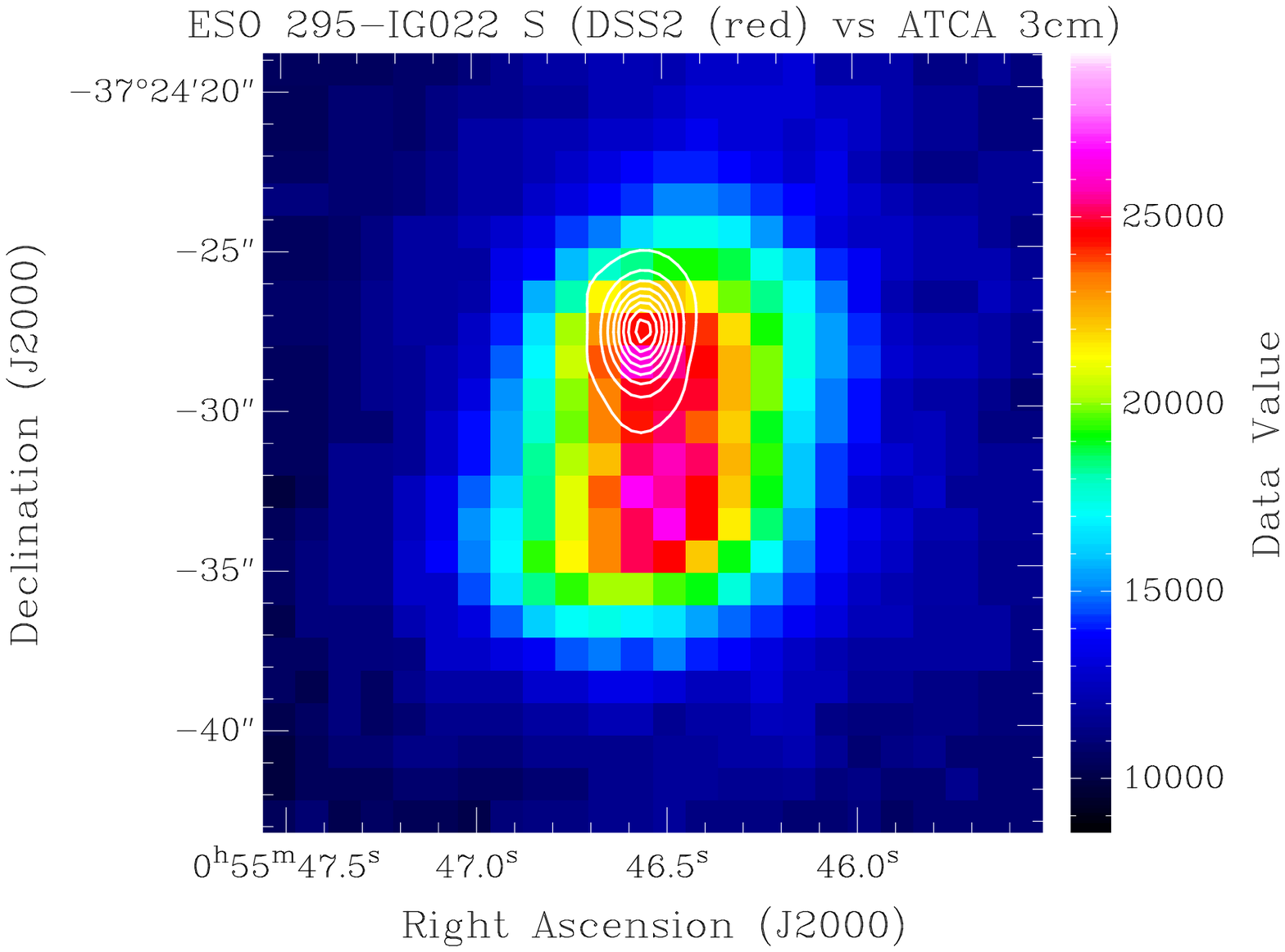}}
\vskip-5mm
\figurecaption{4.}{Digital Sky Survey (red) image of the \ESOs\ region overlaid with primary beam corrected 3-cm ATCA radio-continuum contours. The synthesized beam of the ATCA observation is 2.2$\arcsec$ $\times$ 1.6$\arcsec$ with r.m.s. noise (1$\sigma$) of 0.1~mJy. Contours increase by factors of 50$\sigma$ from 1.5 mJy/beam.}

The linear polarisation image of \AB\ at 6~cm is illustrated in Fig.~6. This linear polarisation image was created using \textit{Q} and \textit{U} parameters. Regions of fractional polarisation are quite strong at the southern side of the \AB\ cluster with polarisation vectors aligned with the observed jets from \ESOn. We could not determine the Faraday rotation and magnetic field without polarisation measurements at the second wavelength of 3~cm. The mean fractional polarisation at 6~cm was calculated using flux density and polarisation:
\begin{equation}
P=\frac{\sqrt{S_{Q}^{2}+S_{U}^{2}}}{S_{I}}\cdot 100\%
\end{equation}
\noindent where $S_{Q}, S_{U}$ and $S_{I}$ are integrated intensities for \textit{Q}, \textit{U} and \textit{I} Stokes parameters. Our estimated peak polarisation value is $P\cong 72\% \pm15\%$ at 6~cm.  Along the northern part of jet there is a pocket of uniform polarisation at approximately 70\% (Fig.~6) possibly indicating varied dynamics along the jet. We speculate that in the mid part of the jet some twisting of the two bended jets may produce poorly ordered polarisation as can be seen also in Fig.~6. This strong and ordered polarisation is among the strongest ever observed in a large scale AGN.

}\end{multicols}

%
%
\centerline{
 \includegraphics[width=.33\textwidth,angle=-90]{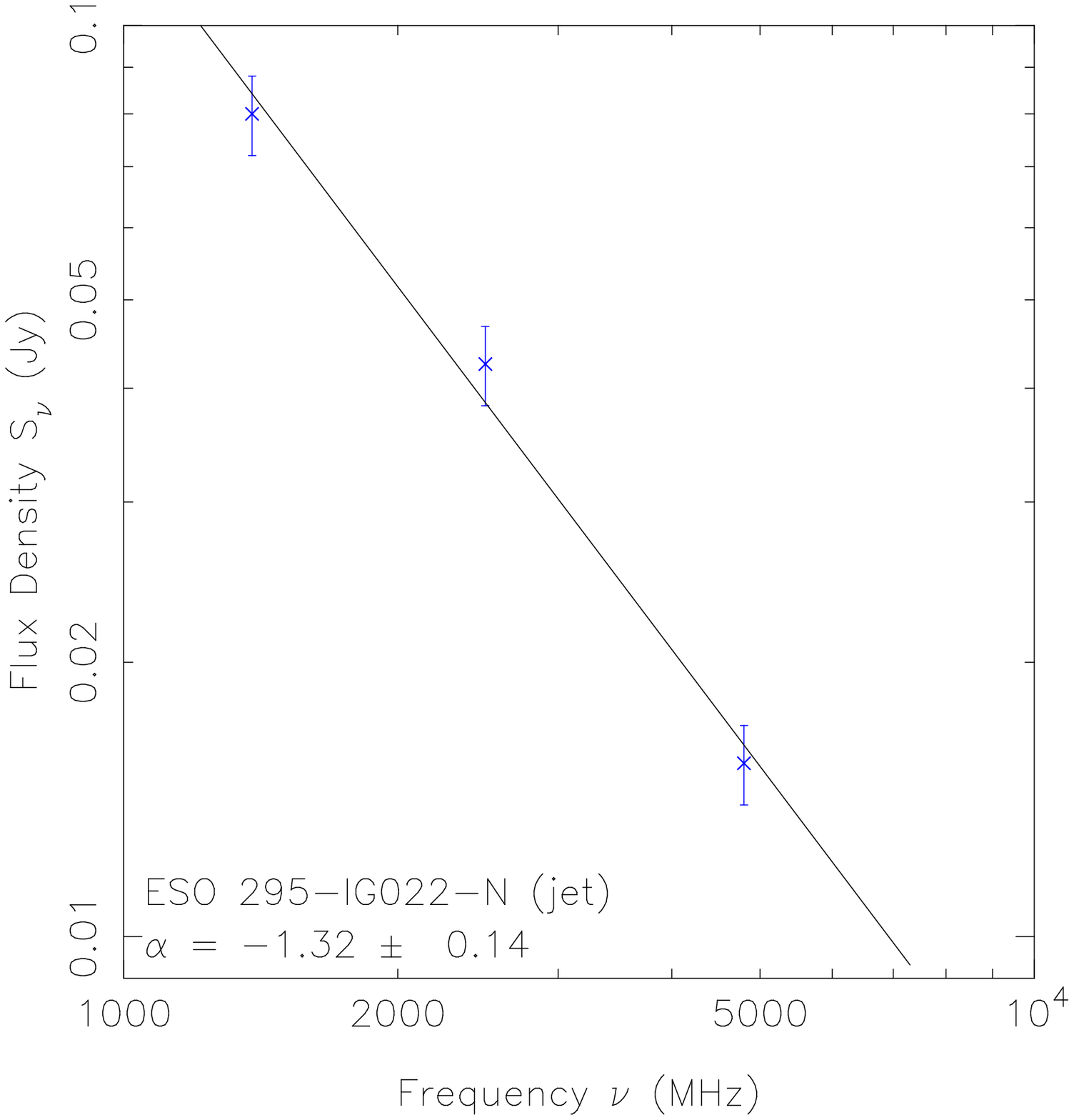}
 \includegraphics[width=.33\textwidth,angle=-90]{Fig-5b.eps}
 \includegraphics[width=.33\textwidth,angle=-90]{Fig-5c.eps}}

\figurecaption{5.}{Spectral index distribution of the radio features within the \AB. (a) Jet structure associated with \ESOn\ based on the ATCA 2010 observations (b) \ESOs\ based on our ATCA 2000 observations (c) \ESOs\ from Burnett \etal\ (2010). For (b) \& (c) $\alpha_T$ is the fit including all points, $\alpha_1$ and $\alpha_2$ are fits using a subset of the points.}

\begin{multicols*}{2}
{

%
%
\vskip3mm
\centerline{\includegraphics[width=8.5cm,angle=-90]{Fig-6.ps}}
\figurecaption{6.}{ATCA observations of \ESOn\ at 6~cm (4.8~GHz). The blue circle in the lower left corner represents the synthesised beamwidth of 4\,\arcsec$\times$3\,\arcsec\ and the blue line below the circle is a polarisation vector of 100\%. }

We note that \ESOn\ is typical of narrow angle tail radio galaxies which are commonly found in rich Abell clusters. While it looks like the emission is weak, the second tail is now seen well. 

\section{4. SUMMARY}

We present new high resolution images at various radio-continuum frequencies of the peculiar galaxy pair \ESO\ at the centre of the poor cluster \AB. The most likely interpretation is that the double jet structure originates from the northern galaxy (\ESOn), which then bends and twistes towards the south. We detected strong polarised emission of $\sim70\%$ from the AGN jets. Also, the spectral index of jet filaments is very steep ($\alpha$=--1.32) indicating the presence of strong magnetic fields. No point source was detected that could be associated with the core of \ESOn. 

On the other hand, \ESOs\ does not show any jet structure, but does have a radio point source with variable flux and spectral index. This appears to be superposed in the line-of-sight of the 20-cm jets, which most likely originate from the northern galaxy \ESOn\. We emphasis that the more sensitive and higher resolution radio-continuum observations of this Abell cluster would help to reveal the real nature of this complex system.


\acknowledgements{We used the {\sc karma/miriad} software package developed by the ATNF. The Australia Telescope Compact Array is part of the Australia Telescope which is funded by the Commonwealth of Australia for operation as a National Facility managed by CSIRO. PAJ acknowledges partial support from Centro de Astrof\'\i sica FONDAP 15010003 and the GEMINI-CONICYT FUND. We thank the referee for numerous helpful comments that have greatly improved the quality of this paper.
}

\newcommand{\MNRAS}{\journal{Mon. Not. R. Astron. Soc.}}
\newcommand{\ApJ}{\journal{Astrophys. J.}}
\newcommand{\ApJS}{\journal{Astrophys. J. Supplement}}
\newcommand{\AJ}{\journal{Astronomical. J.}}

\references

Abell, G. O., Corwin, H. G., Olowin, R. P.: 1989, \ApJS, \vol{70}, 1.

Bica, E. L. D., Pastoriza, M. G., Maia, M., da Silva, L. A. L., Dottori, H.: 1991, \AJ, \vol{102}, 1702.

B\^irzan, L., McNamara, B. R., Carilli, C. L., Nulsen, P. E. J., Wise, M. W.: 2007, in: Heating versus Cooling in Galaxies and Clusters of Galaxies, ESO Astrophysics Symposia. Springer-Verlag Berlin Heidelberg, p. 115.

Burnett, C., Lonsdale, N., Pearce, K.: 2010, ATNF Summer Student Symposium, http://www.atnf.csiro.au/internal/meetings/2010/0008.html

Douglass, E. M., Blanton, E. L., Clarke, T. E., Sarazin, C. L., Wise, M. W.: 2008, \ApJ, \vol{673}, 763.

Gooch, R.: 2006, Karma Users Manual, Australia Telescope National Facility.

Jones, P. A., McAdam, W. B., Reynolds, J. E.: 1994, \MNRAS, \vol{268}, 602.

Lumsden, S. L., Nichol, R. C., Collins, C. A., Guzzo, L.: 1992, \MNRAS, \vol{258}, 1. 

McAdam, W. B., White, G. L., Bunton, J. D.: 1988, \MNRAS, \vol{235}, 425. 

Payne, J. L., Filipovi\'c, M. D., Pannuti, T. G., Jones, P. A., Duric, N., White, G. L., Carpano, S.: 2004, \journal{Astron. Astrophys}, \vol{425}, 443.

Rafferty, D. A., McNamara, B. R., Nulsen, P. E. J., Wise, M. W.: 2006, \ApJ, \vol{652}, 216.

Read, A. M., Filipovi\'c, M. D., Pietsch, W., Jones, P. A.: 2001, \journal{Astron. Astrophys}, \vol{369}, 467.

Sanders, J. S., Fabian, A. C., Taylor, G. B.: 2009, \MNRAS, \vol{393}, 71.

Sault, B., Killeen, N.: 2010, {\sc miriad} users Guide, ATNF.

Wise, M. W., McNamara, B. R., Nulsen, P. E. J., Houck, J. C., David, L. P.: 2007, \ApJ, \vol{659}, 1153.

\endreferences

}
\end{multicols*}

\vfill\eject

{\ }


\naslov{RADIO KONTINUM MLAZEVI U OKOLINI peculiar PARA GALAKSIJA \mbox{\rm\bf ESO\,295-IG022}}


\authors{M.~D.~Filipovi\'c$^1$, E.~J.~Crawford$^1$, P.~A.~Jones$^{2,3}$ and G.~L.~White$^1$}

\vskip3mm


\address{$^1$University of Western Sydney, Locked Bag 1797, Penrith South DC, NSW 1797, Australia}
\address{$^2$School of Physics, University of New South Wales, Sydney, NSW 2052, Australia}
\address{$^3$Departamento de Astronom\'\i a, Universidad de Chile, Casilla 36-D, Santiago, Chile}

\vskip3mm


\centerline{\rrm UDK \udc}

\vskip1mm

\centerline{\rit Originalni nauqni rad}

\vskip.7cm

\begin{multicols}{2}

{


\rrm 

U ovoj studiji predstav{lj}amo nove {\rm ATCA} radio-kontinum rezultate posmatra{nj}a regiona u okru{\zz}e{nj}u neobiqnog para galaksija \textrm{\ESO} u centru slabog jata \textrm{\AB}. 
Posmatra{nj}a su vrxena sa 6~km konfiguracijom na talasnim du{\zz}inama od \textrm{$\lambda$=20/13} i 6/3~cm. Rezolucija je $\sim$2\arcsec\ na 3~cm xto je dovo{lj}no za razluqiva{nj}e jet-like strukture od $\sim$3\arcmin\ du{\zz}ine orginalno detektovane na 20~cm. Koriste{\cc}i naxa nova radio posmatra{nj}a potvr{dj}ujemo postoja{nj}e double jet strukture koje je nastala iz severnije galaksije \textrm{(\ESOn)}. Oba mlaza su savijena i uvijena prema ju{\zz}noj galaksiji \textrm{(\ESOs)}. Radio spektralni indeks mlaza nastalog iz severne galaksije \textrm{(\ESOn)} je veoma strm \textrm{($\alpha$=--1.32)} xto bez obzira na ne postoja{nj}e centralnog taqkastog objekta potvr{dj}uje izvornost Aktivnog Galaktiqkog Jezgra (AGJ). Sa druge strane, \textrm{(\ESOs)} je taqkasti radio objekat te sa promen{lj}ivim fluksom i spektralnim indeksom ukazuje na centar AGJ koji je projektovan na liniju posmatra{nj}a mlazeva (vi{dj}enih na 20~cm) nastalih iz severne galaksije \textrm{\ESOn}. Detektovali smo pravilno organizovan i veoma visok stepen polarizacije gde je maksimalna vrednost iznosi oko 70\%. 

}
\end{multicols}

\end{document}